\documentclass[mathptmx
    ,final            
  ]
  {aipproc}

\layoutstyle{8x11double}

\def\gx{GX~339$-$4}
\def\cx{Cyg~X$-$1}

\def\grs{GRS~1915$+$105}
\def\gro{GRO~J1655$-$40}

\def\grsds{GRS~1758$-$258}

\def\1e{1E~1740.7$-$2942}

\def\cir{Cir~X$-$1}
\def\xte{XTE~J1550$-$564}
\def\xt{XTE~J1650$-$500}
\def\xteonze{XTE~J1118$+$480}
\def\xtedixsept{XTE~J1748$-$288}                   
\def\ds{4U~1755$-$33}

\newcommand{\sm}      {\mbox{$\rm\,M_{\mathord\odot}$}}

\def\arcmin{\hbox{$^\prime$}}
\def\arcsec{\hbox{$^{\prime\prime}$}}

\newcommand\rxte{\textsl{RXTE}}    

\begin{document}

\title{Relativistic Jets in the {\em RXTE } Era}

\author{Stéphane Corbel }{
  address={Universit\'e Paris 7 Denis Diderot and Service d'Astrophysique,
                CEA Saclay, F-91191 Gif sur Yvette, France }
}

\begin{abstract}

Since the launch of the Rossi X-ray Timing Explorer in 1995
our understanding of jetted outflows has significantly improved.
Indeed, relativistic jets are now believed to be a fairly ubiquitous
property of accreting compact objects, that are intimately coupled with
the accretion history. In this review, we summarize the observational
connections  in X-ray binaries between accretion flows and relativistic
outflows (especially the relation with the X-ray states). 
We emphasize those aspects that have significantly benefited from the RXTE 
experiment, including the role that jets could play at high energies. 
We also review recent observations of large scale relativistic jets 
that could point to their long term effects on the interstellar medium. 
\end{abstract}

\maketitle


\section{Black hole candidates and X-ray states}

Black hole candidates (hereafter BH, even if its mass function has not been measured)
 are known to exhibit several X-ray spectral
states, distinguished by the presence or absence of a soft black-body component at
$\sim$ 1 keV (arising from the accretion disk) and the luminosity and spectral slope 
of emission at harder energies (whose nature is still the subject of an active debate).
Systems in the low-hard state (LHS) have power-law X-ray spectra with a photon index
in the range 1.4--1.9, an exponential cutoff (but not always) around 100 keV, and no (or only weak) 
evidence of a soft thermal component. In the LHS, the power density spectrum (PDS) is characterized 
by a strong band-limited noise power continuum. At higher soft X-ray flux\footnote{However, the 
LHS is sometimes observed at higher flux \cite{tom03a}.}, these systems are usually found in 
the thermal-dominant (TD) or high-soft (HS) state. In that case, the X-ray spectrum is 
dominated (up to $\sim$ 90\%) by the soft-thermal component, with an
additional weak and steep power-law component (with no apparent cut-off) at 
higher energy. Very little variability is detected in the PDS of BHs while in the TD state.
At much higher flux, in the steep power-law (SPL) or very high state (VHS), both the disk
black-body component and the steep power-law component at higher energy are detected with relative
contributions that can vary significantly. The so-called intermediate state (IS) may be very 
similar to the VHS \cite{hom01}, but it occurs at lower luminosity \cite[see also][]{mcc04}. 
Outside the bursting period, these systems are usually found 
in a quiescence state (QS) with very weak residual X-ray emission that displays many similarities
with the LHS \cite[e.g. ][]{cor00,tom04}. This implies that the QS could simply be a low luminosity version 
of the LHS. With the detailed coverage by RXTE of spectral properties of BHs during recent outbursts, 
the different X-ray states can be well illustrated using a Hardness-Intensity diagram (see Fig. 1
from \cite{bel03}). In this case, the VHS and IS are defined as various tracks between the 
vertical tracks (of constant hardness) of the HS and LHS \cite{hom01,bel03}. For further details on the 
spectral states of BH, see \cite{mcc04}. In the sections that follow, we  discuss the general properties of BHs
in the various X-ray states, some of which they have in common with neutron star (NS) systems, and focus on
relativistic jets in the light of recent developments based on multi-wavelength observations (including RXTE).
For a detailed reviews of jets in X-ray binaries, see \cite{fen04a}.

\begin{figure}[hbt]
  \includegraphics[height=.25\textheight]{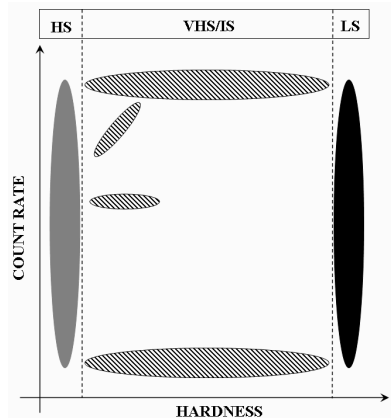}
  \caption{Schematic Hardness-Intensity diagram (HID), from \cite[]{bel03}, illustrating the
location of the different X-ray states in the HID. }
\end{figure}

\section{Transient jet ejection events}

\subsection{From ``Superluminal'' Ejections ... }

During period of outburst activity in BHs, strong radio flares\footnote{For a reasonable number of sources, 
only a weak (few mJy) radio flare is detected.} are sometimes observed  around the
transition from the hard state (QS or LHS) to a soft state (TD or SPL state). This is usually interpreted 
as synchrotron emission from relativistic electrons ejected from the system with large bulk velocities. 
In a few cases such jets have been directly imaged into one-sided 
(or two-sided) components moving away from the stationary core. In those observations, motions were detected 
with apparent velocities greater than the speed of light (very similar to phenomenon observed in Active 
Galactic Nuclei). 
The inferred bulk Lorentz factor\footnote{$ \Gamma = 1/\sqrt{1 - \beta^2}$, with the velocity v defined 
in unit of c as $v = \beta c$.} is typically in the range 2 to 5 \cite{fen04a}. After ejection, the moving 
plasma condensations are observed in the radio range for a few weeks until their emission fades below detection levels 
due to their expansion. Prior to the launch of \rxte, only two microquasars were known among the population of 
BHs: \grs\ (see Fig. 2 from \cite{mir94}) and \gro\ \cite{tin95,hje95}. 

\begin{figure}[hbt]
 \includegraphics[height=.35\textheight]{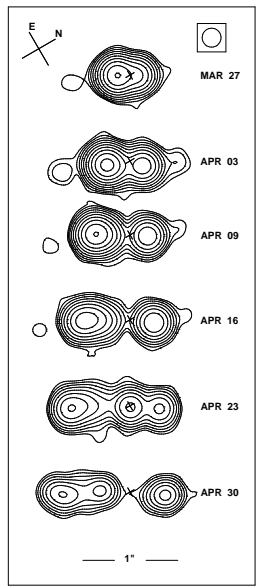}
 \caption{VLA 8.4 GHz maps illustrating  the discovery \cite{mir94} of relativistic ejections from GRS 1915+105 in 1994. 
 Reprinted by permission from Nature (Mirabel \& Rodr\'\i guez, Nature, 371, 46) copyright (1994) Macmillan Publishers Ltd.}
\end{figure}

Since then, the number of BHs displaying apparent superluminal motion has greatly increased. For example \xte\ 
\cite{han01}, \xtedixsept\ \cite{rup98}, V4641~Sgr \cite{hje00}  exhibited such behaviours and it is reasonable to think that all 
BHs (and also some NS) are likely to exhibit highly relativistic jets at some time in their history 
(see for example 4U~1755$-$33 \cite{ang03} or \gx\ \cite{gal04}). In fact, in recent years, almost 
all active BHs have been associated with radio emission. And this is not a unique 
``privilege'' of BH, as some NS systems are associated with highly relativistic outflow (e.g.  
Sco~X-1 \cite{fom01} and \cir\ \cite{fen04b}). In those cases, the knots moving with mildly relativistic 
bulk velocity (from 0.01 to 0.5 c) are energized by an unseen beam of particles that could be
ultra-relativistic with bulk Lorentz factor $>$ 10 for \cir\ \cite{fen04b} (or $>$ 3 for  Sco~X1, 
\cite{fom01}). In this different scenario, the observed synchrotron emission would be powered locally 
during  the interaction of the unseen beam with the interstellar medium (ISM) or by moving shocks 
within the flow itself. 

\subsection{... to Large Scale Jets}

Beside the sub-arcsecond scale transient jets seen during the initial part of the outburst of some 
BHs, large scale (few arc-minutes) stationary radio jets have been observed in a few systems. 
Indeed,  the BHs \1e\ (Fig. 3 from \cite{mir92}) or \grsds\ \cite{mar02} in the Galactic Bulge are 
located at the center of two large scale radio lobes, probably indicating the long term action of 
past ejections on the surroundings ISM.  

\begin{figure}[hbt]
  \includegraphics[height=.3\textheight]{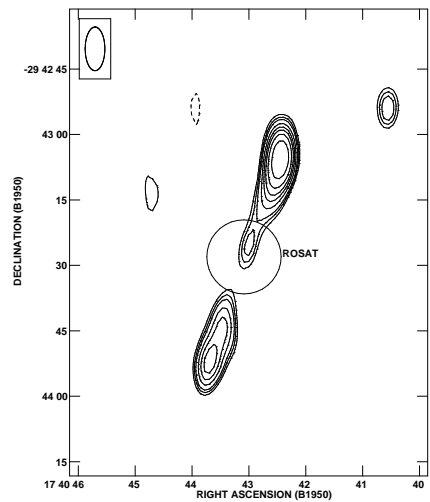}
  \caption{Large scale radio jets (at 4.8 GHz) from 1E1740.7-2942. Reprinted, with permission, from the Annual Review 
 of Astronomy and Astrophysics, Volume 37, Copyright by Annual Reviews www.annualreviews.org. }
\end{figure}

Recent X-ray observations (Fig. 4) by Chandra   have led to
the discovery of extended (up to 30\arcsec) X-ray jet emission from the microquasar \xte. 
In observations made between June 2000 and January 2003, two sources moving away 
from the \xte\ black hole are detected. 
Both the radio and X-ray emission of the western jet appeared extended
towards \xte, and the morphologies associated with each wavelength matched well.
The broadband spectra of the jets are consistent
with synchrotron emission from high-energy (up to 10~TeV)
particles accelerated in shocks formed by the interaction of the jets
with the ISM \cite{cor02a} (i.e. similar to the stationary non-thermal
emission from the large scale lobes in SS~433 \cite{sew80}). 

\begin{figure}[hbt]
  \includegraphics[height=.5\textheight]{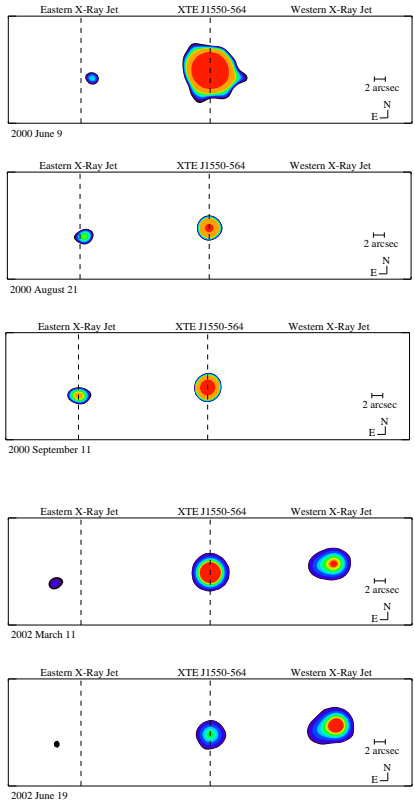}
  \caption{Five {\em Chandra} 0.3-8~keV images showing
XTE~J1550--564, the evolution of the eastern and western  X-ray jets
between June 2000 and June 2002.
The observations are ordered chronologically from top to
bottom, and each image is labeled with the observation date.
The dashed lines mark the positions of XTE J1550-564 and the eastern
X-ray jet on 11 September 2000.  Adapted from \cite{cor02a,tom03b,kaa03}. Reprinted from New Astronomy Reviews, Vol 47,
     2003, Corbel et al., p 477, Copyright (2003), with permission from Elsevier.  }
\end{figure}

The most likely scenario \cite{cor02a} is that the eastern jet 
is the approaching jet and the western jet the receding jet, and 
that the jet material was ejected
from the black hole during the major radio flare of September 1998
\cite{han01}.
The full set of X-ray and radio observations also provided the first direct
evidence for gradual  deceleration of relativistic materials in a jet.
Previous observations of other microquasars were consistent with purely
ballistic motions except for \xtedixsept, where, after ballistic
ejections, the jet was observed to stop over the course of a few weeks. 
presumably following a collision with environmental material \cite{kot00}.
More details on the X-ray jets of \xte\ can be found in \cite{cor02a,tom03b,kaa03}.
These results indicate that emission due to the relativistic plasma
ejected in September 1998 has been detected for at least 5 years (Corbel, 
private communication) as direct beamed X-ray emission, and demonstrate 
that Galactic BHs are able to accelerate particles up 
to very high energies.   

\gx\ is one of the best-studied black hole binaries, and has
been the key source (see below) for unraveling the association of X-ray states and
the formation of jets in accreting black holes \cite{fen99a,cor00,cor03,gal04}.
Following more than two years in quiescence, the source re-brightened in early 
2002 to its brightest level in a decade.
In May 2002, while brightening in X-rays, \gx\ produced an intense and
rapid radio flare  (associated with a dramatic X-ray state
change from a LHS to SPL state) \cite{gal04}.
The peak flux of the flare was $>50$ mJy, making it the
brightest level of radio emission ever observed in \gx. Further observations with ATCA
have tracked the formation of a well-collimated one-sided jet (Fig. 5) extending to about 12 arcsec,
with apparent velocity greater than 0.9c \cite{gal04}. This jet is consistent  with shocks 
waves formed within the jet itself (as several radio flares were observed) and/or by the action 
of an underlying highly relativistic outflow with the ISM \cite{gal04}. 

\begin{figure}[hbt]
  \includegraphics[height=.35\textheight,angle=270]{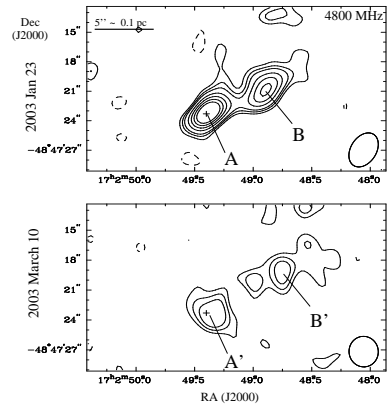}
  \caption{4.8 GHz maps of the extended jet of \gx.  Top: 2003 January 23;
	bottom: : 2003 March 10. Figure from \cite{gal04}.}
\end{figure}

These results are remarkable and reminiscent of the decelerating jets of \xte. But in the case
of \gx, the luminosity of the jets decreased much more rapidly by being undetectable at
radio frequencies in less than a year \cite{gal04}.  
With the detection of this relativistic outflow in a recurrent BH,  \gx\ could  now 
be classified as a microquasar. In fact, we note that all soft X-ray transient are also potentially 
microquasars; the superluminal ejections could be missed due to their short lifetime (if the jets 
are almost aligned with the observers) and/or the lack of radio observation. 
The infrared jet observed only once in \grs\ may have the same origin as in \xte\ cite{sam96}. 

Very recently, Angelini \& White \cite{ang03} reported the 2000 XMM-Newton detection 
of large scale (7\arcmin) persistent X-ray jets (Fig. 6)
centered on the position of \ds, a black hole candidate.  \ds\
had been a bright, persistent source for at least twenty years until it
became X-ray quiet in 1995.  If the jet had a velocity close to $c$,
then it would have taken about 13 years to expand to its currently
observed length of $\approx 4$~pc (for a distance of 4~kpc).  The
jet/ISM interaction in \ds\ might be similar to that seen in \xte ,
provided the jets were being ejected quasi-continuously over its twenty
years X-ray activity.

\begin{figure}[hbt]
  \includegraphics[height=.35\textheight, angle = 270]{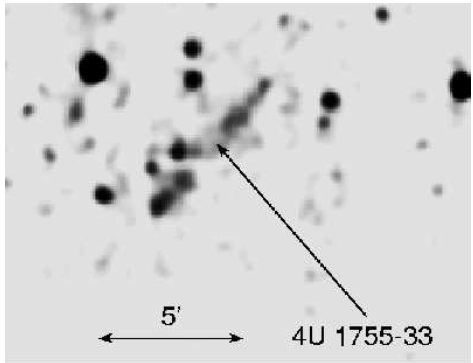}
  \caption{{\em XMM-Newton} image of \ds\ \cite{ang03}. The arrow indicates the position
of \ds. Figure adapted by P. Kaaret \cite{kaa03b} from \cite{ang03}.}
\end{figure}

The scale of the moving X-ray and radio lobes (0.5 to 0.8 pc) 
in \xte\ is intermediate in size between the moving ``superluminal'' 
ejections very close ($<$ 0.1 pc) to the compact object (like in e.g. \grs) and
the stationary lobes (1-4 pc) such as in \1e\ or in \ds. This would 
suggest a morphological evolution: in that case, the large scale stationary 
lobes would be the results of the long term action of past relativistic ejections
on the local ISM.

\section{Ubiquitous Powerful Compact Jets Associated to the Low-Hard State}

In addition to the more ``spectacular superluminal'' ejections from these systems, 
a new kind of jet has been introduced these recent years that might have
much more impact on the spectral energy distribution (SED) of BHs (especially in the LHS).  
RXTE has been of great importance for demonstrating the strong coupling between
such jets and the accretion properties.

\subsection{From Radio Emission ...}

Indeed, the LHS has now been observed very frequently at radio frequencies and a radio 
source is almost always detected with a flat or slightly inverted
spectrum (spectral index $\alpha \ge 0$ for a flux density S$_\nu$ $\propto$ $\nu^\alpha$)
and a level of linear polarization of $\sim$ 1--2\% \cite[e.g.][]{mar96,cor00,fen01}. In addition, 
the radio counterpart of \cx\ has been directly resolved into a few milliarcsecond-scale 
outflow during its standard LHS (Fig. 7) \cite{sti01}. These properties are
characteristic of a conical self-absorbed compact jet, similar to that considered
for flat spectrum AGNs \cite{bla79}. 
Similar properties have now been observed in a growing number of persistent
and transient BHs, thus suggesting that compact jets are ubiquitous
in BHs during the LHS \cite{fen01}.

\begin{figure}[hbt]
  \includegraphics[height=.35\textheight,angle=270]{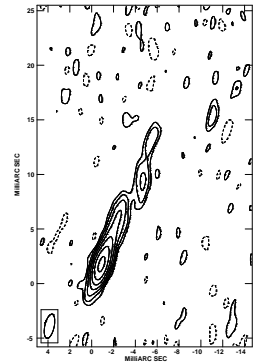}
  \caption{One-sided self absorbed compact jet from \cx\
in the LHS \cite{sti01} with the VLBA at 8.4 GHz. }
\end{figure}

\subsection{... to Near Infrared ...}

In addition to being responsible for most of the emission in the radio regime, 
the compact jets may have a significant contribution in the infrared/optical bands.
In some BHs, the infrared flux densities are consistent with an extension of the inverted 
radio spectral component \cite{fen01,cor01}.
Furthermore, Jain et al. \cite{jai01} observed a secondary flare, prominent in near infrared, 
but also in optical, in the light-curve of \xte\ during a transition (in 2000) to the LHS,
and therefore associated  with the reappearance of the compact jet \cite{cor01}. This has now also been
observed in \gx\ and 4U~1543$-$47 and in fact this may be a common occurrence during outburst
decline \cite[ in these Proc.]{bux03}. Infrared synchrotron emission has already been invoked
to explain the correlated radio-infrared flares of \grs\ \cite{fen97, mir98, eik98}.

So if the self-absorbed synchrotron spectrum extends from radio to near-infrared frequencies, then
above some frequency there should be a break (as observed in AGNs) frequency above which the
spectrum of the jet is no longer self-absorbed, even at its base. Corbel \& Fender \cite{cor02b} have
identified this cut-off frequency (Fig. 8) in the near-infrared observations of \gx\
during a bright LHS.  In that case, the power in the jet would be at least greater than 10\%
of the X-ray luminosity \cite{fen01,cor01, cor02b}.

\begin{figure}[hbt]
  \includegraphics[height=.35\textheight,angle= 270]{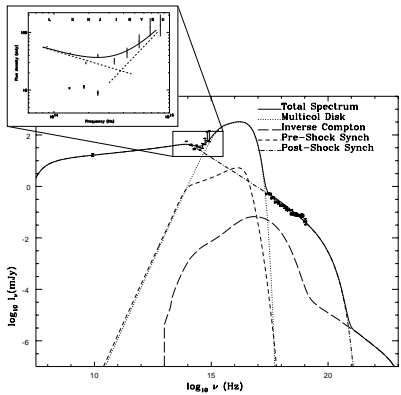}
  \caption{Broadband spectrum of \gx\ in the LHS, the continuous line
represents the fit (with the various components) with the jet/disk model
of Markoff et al. \cite{mar01,mar03}. The insert (from \cite{cor02b}) illustrates 
the turnover frequency in near infrared, above which the compact jet is
no longer self-absorbed.} 
\end{figure}

\subsection{ ... up to X-ray Jet Emission}

Such evidence for synchrotron emission (from the compact jets) extending to -at least-
near infrared naturally raises the possibility of even higher energy (i.e. X--ray) synchrotron
emission from optically thin emission above the turnover. And indeed, extrapolation of 
the optically thin synchrotron component from the near-infrared to higher frequencies
sometimes coincides with the observed X-ray spectrum \cite{cor02b}, supporting models in which
the X-rays could originate via optically thin synchrotron emission from the jet 
(possibly instead or in addition of Comptonisation) \cite{mar01}.

As a result from a long-term study of the BH \gx\ using simultaneous radio 
and X-ray (mainly  RXTE) observations performed between 1997 and 2000,  
Corbel et al. \cite{cor00, cor03} have found a strong correlation (indicating a strong 
coupling) between these emission
regimes extending over more than three  orders of magnitude in soft X-ray flux.
The radio emission scales as $F_{\rm radio} \propto F_{\rm X-ray}^b$ where 
$b \sim 0.7$ for X--rays up to at least 20 keV. 

\begin{figure}[hbt]
  \includegraphics[height=.35\textheight, angle = 270]{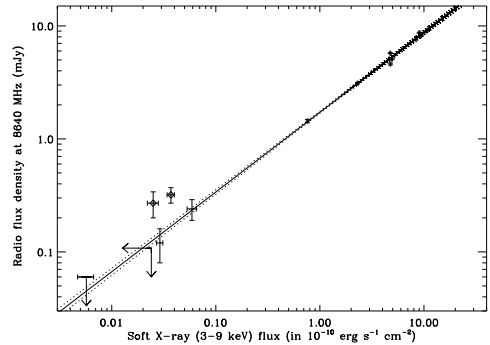}
  \caption{\gx\ (1997-2000): Radio flux density at 8.6 GHz versus the
   X-ray flux in the 3-9 keV energy band (from \cite{cor03}).}
\end{figure}

It is also interesting to note that Gallo, Fender \& Pooley \cite{gal03} have found
a similar coupling (with the same normalization and slope) for the transient
V404~Cyg. Furthermore, they found using data for ten other BHs in the LHS
that they were all consistent with the same coupling.  The small scatter 
would imply low bulk Lorentz factor for the compact jets ($\Gamma < 2$)
\cite{gal03}.

Similarly,  an almost identical (i.e lying on the extrapolation obtained by \cite{cor03, gal03}) 
non-linear coupling is observed in a large sample of supermassive black holes, if 
one takes into account the mass of the black hole as an additional (and natural) term 
in the  scaling relation between radio and X--ray emission \cite{fal03,mer03}. This 
would suggest that the same physics applies over many orders of magnitude in 
black hole mass and  that the spectral energy distribution (SED) of BHs operating
at sub-Eddington accretion rate could be dominated by non-thermal emission from a
relativistic jet \cite{fal03}.

As an additional remark, we should point out the recent results by Hynes et al. \cite{hyn03} 
which demonstrate that the broadband variability spectrum (from infrared to X--ray) 
of \xteonze\ were fully consistent with expectations of optically thin synchrotron emission.

\subsection{Broadband Jet Emission}

In the past years, a lot of effort has been made in organizing campaigns of  multi-wavelength
observations simultaneously with RXTE. This allowed the 
collection of a large number  of SEDs (obtained at many different flux levels)
that need to be modelized carefully in order to shed some light on the physical 
processes at work in these systems.  Such attempt has been originally made by Markoff, Falcke
and Fender \cite{mar01} in the case of \xteonze\ using a jet-disk model (details can be found in 
\cite{mar01, mar03}). Using thirteen broadband spectra of \gx\ over at least two orders of
magnitude in X--ray flux, Markoff et al. \cite{mar03} have shown that this
jet-disk model qualitatively agrees with all SEDs (e.g. Fig 8) if one changes only
two parameters in the model: the input power and the location of the
first acceleration zone.  It is interesting to note that the broadband 
spectra and model fits are comparable to those for XTE J1118+480 while in the same X-ray
state.  Furthermore, this model can analytically explain
the slope (Fig. 9) of the observed radio/X-ray correlation \cite{cor03}
by changing only the input power in the jets. 
The implication of the model, if correct, would be that the majority of
power output in the low/hard state is dissipated via the jets.  Jet
synchrotron is therefore a natural way to explain the broadband
features and this correlation. 

The high energy contribution of the jets in the SED could take a different form  as in the case of 
comptonization by jet electrons of external photons from the companion star \cite{geo02}.
The correlation between radio and X-ray emission may also be
interpreted within the framework of the Two Component Advective Flow
(TCAF) model \cite{cha96}, in which the location of a boundary
layer between the thin disk and the Comptonizing region determines the
spectral shape and also the amount of outflow \cite{cho03}.

These recent studies emphasize the importance of taking the existence of
jets into account when modeling the SED of BH (see also the review in \cite{mcc04} for more details
on other plausible interpretations of the high energy contribution), even if Comptonizing corona and 
synchrotron emitting jets are not necessarily mutually exclusive (e.g. corona = base of the jets).
Such hybrid models (including also the reflection component) should now be considered in more details
in order to quantify the total contribution of the jets in the SED of BHCs.

\subsection{Quiescence State}

The observed radio/X-ray flux correlation in \gx\ and V404~Cyg seems to extend down
to almost their quiescence level, i.e. down to $10^{-6}$ times the Eddington luminosity 
(for a 10 \sm\ BH) \cite{cor03,gal03}. This could indicate that in terms of radio
properties, the QS could be very similar to the LHS as already suggested by 
some X--ray properties  \cite{cor00,tom03a}. Furthermore, at even lower luminosity,
the BHs could be in a jet-dominated state in which the liberated 
accretion power would be in the form of a radiatively inefficient outflow and not 
dissipated in the accretion flow \cite{fgj04}. This may be sufficient to explain the luminosity difference
between BHs and NSs in QS \cite{fgj04}.

\section{Quenched radio emission in the TD state}

On the other side of the  radio/X-ray flux correlation curve, i.e. at high X-ray flux, it is found that 
above a few percent of Eddington luminosity, the radio emission is gradually suppressed \cite{gal03}.
The quenching of radio emission is well illustrated in Fig. 10 showing the 1998 TD state of \gx\ 
\cite{fen99a,cor00}.  This phenomena was in fact already observed in \cx\  with the reappearance of radio
emission during a transition in 1996 from TD (or IS, see \cite{bel99}) to a  LHS \cite{zha97} 
(see also \cite{tan72}) and has been 
found again during a TD state in 2002 \cite{tig04}. In addition, a large sample of AGN shows similar 
quenching when the luminosity is above a few percent of the Eddington luminosity \cite{mac04}. This would  again 
suggest that the same disc-jet coupling exists in these two types of BHs and thus that the same physics
regulates these different systems.

\begin{figure}[hbt]
 \includegraphics[height=.35\textheight, angle = 270]{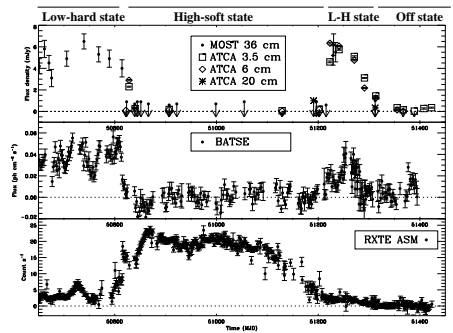}
  \caption{Radio, soft and hard X-ray emission from \gx\ between summers 1997 and 1999. The transition
to a TD state (or HS) is associated with quenched radio emission \cite{fen99a,cor00}.}
\end{figure}

The situation could be more complicated if we take into account the detection of radio emission
during the TD state of \xt\ \cite{cor04} and  \gx\ \cite{gal04} in their recent outburst. However, 
their observed  properties are drastically different 
from those in the LHS especially with an optically thin radio spectrum.  For \gx, this has clearly been 
associated with the formation of a large scale jet (possibly during a collision of relativistic plasma 
with the ISM) \cite{gal03}. This explanation is also plausible in the case of \xt\ \cite{cor04}, which 
would suggest that the interaction of the jets with the ISM (as in the case of \xte, \cite{cor02a}) is
 more common than previously thought.

\section{The SPL state}

The SPL state, originally believed to be very rare, is now know to appear in almost all BHs
during outburst, thanks to RXTE observations \cite{mcc04}.
However, this state has been very rarely observed
at radio frequencies. When observed, the radio emission were usually
dominated by the decaying optically thin synchrotron emission arising
from the radio flare associated to the quiescent/active state transition
and therefore decoupled from the BH system.  During an outburst 
in 2000, \xte\ made a transition from a LHS to a SPL state (the transition
was associated with a small radio flare) \cite{cor01}. A few days later, the
following radio observation was not detected \xte, indicating
a quenching of radio emission by a factor
$>50$ compared to the initial LHS \cite{cor01}.  Furthermore, the state in which quenched radio emission of \cx\
was observed in 1996 may also be the IS (an instance of the SPL state ?)  and
not the TD state \cite{bel99}. So these two sources would suggest that the SPL
state is associated with quenched radio emission. 

However, the behaviour of \grs\ may suggest a slightly different pattern. 
The comparison of \grs\ with ``canonical'' BH states is not straightforward, but 
the X-ray states A, B (soft) and C (hard) may be similar to the SPL state
\cite{rei03}. Oscillations between states A, B and C are correlated
with radio flaring activity  (e.g. \cite{kle02}) with stronger
radio emission in state C (associated with spectral hardening), whereas
the soft states are never associated with bright radio emission \cite{kle02}.

As seen previously, \gx\ made a transition from a LHS to a SPL in 2002 \cite{gal04}, 
which was associated with a bright radio flare. Following this flare (while \gx\ was 
still in the SPL state),  the radio behaviour of \gx\ is consistent with multiple 
ejection events, but also with period (at least one) of quenched radio emission \cite{gal04}.

Similarly, \xt\ made a transition from a LHS to a SPL state in 2001. As is seen in \cite{hom02,ros03},
it seems that the SPL state can be decomposed in two parts (groups II and III in \cite{hom02,ros03})
with a gradual softening in group II and a sort of flaring behaviour in group III,
possibly indicating that the accretion disk reached its innermost stable circular orbit in group III.  
Two radio observations have been performed while \xt\ was in the SPL (group III) resulting 
in a weak detection and a non detection \cite{cor04}. This would suggest that the SPL (at least group III) state
would be associated with periods of ejection events (possibly originating from the inner 
accretion disk) interleaved with period of quenched radio emission. 

\section{SUMMARY}

Finally, all these observations performed at many wavelengths have
allowed the investigation of properties
of black holes in their various X-ray states. During the lifetime of RXTE, a rich phenomenology
of behaviour has been brought to light, providing new insights into the accretion-ejection coupling 
in accreting compact objects such as stellar mass BHs and NS, but also supermassive BHs. These can be
summarized as followed:
\begin{itemize}
\item Low-Hard and Quiescence states: always associated with the formation of a compact jet, whose emission
may dominate the entire SED. There are growing indications that the role of the compact jet at high energy 
is significant, but its exact contribution still needs to be quantified precisely.
\item Thermal Dominant state: quenched radio emission whenever the X-ray spectrum is strongly dominated 
by the soft component (from the accretion disk).
\item Step Power Law state: possibly associated with (small scale ?) multiple ejection events. However, 
further observations are needed to constrain ejection properties in this state.
\item From state transition ... to large scale lobes: the transition from a hard state (LHS or QS)  
to a soft state (TD or SPL state) seems to be associated with the ejection of relativistic plasma (that may 
display ``superluminal'' motion). While acting on the ISM (or by internal shocks), repeatedly or not, 
ejection(s) may be responsible for the formation of large scale lobes, with in-situ particle acceleration 
(up to at least TeV energy) and re-energization of the jets very far from the compact object.
\end{itemize}


\begin{theacknowledgments}

I would like to acknowledge Rob Fender, Phil Kaaret, Sera Markoff, Mike Nowak,  J\'erome Rodriguez,
John Tomsick and Tasso Tzioumis for stimulating discussions and for their help in the
various campaigns of multi-wavelength observations. I would also 
like to thank Guillaume Belanger for a careful reading of this manuscript.
 
\end{theacknowledgments}


\bibliographystyle{aipproc}   


\IfFileExists{\jobname.bbl}{}
 {\typeout{}
  \typeout{******************************************}
  \typeout{** Please run "bibtex \jobname" to optain}
  \typeout{** the bibliography and then re-run LaTeX}
  \typeout{** twice to fix the references!}
  \typeout{******************************************}
  \typeout{}
 }

\end{document}